\documentclass{iau379}

\usepackage{amsmath}
\usepackage{graphicx}
\usepackage{multirow}
\begin{document}

\lefttitle{Sedain \& Kacharov}
\righttitle{IAU Symposium 379}

\jnlPage{1}{7}
\jnlDoiYr{2023}
\doival{10.1017/xxxxx}

\aopheadtitle{Proceedings of IAU Symposium 379}
\editors{P. Bonifacio,  M.-R. Cioni \& F. Hammer, eds.}

\title{Testing Jeans dynamical models with prolate rotation on a cosmologically simulated dwarf galaxy}

\author{Amrit Sedain$^{1,2}$, Nikolay Kacharov$^1$}
\affiliation{$^1$ Leibniz Institute for Astrophysics, An der Sternwarte 16, 14482 Potsdam, Germany\\
$^2$ Institute of Physics and Astronomy, Karl-Liebknecht Str.24/25 14476 Potsdam, Germany}

\begin{abstract}
Prolate rotation is characterized by a significant stellar rotation around a galaxy's major axis, which contrasts with the more common oblate rotation. Prolate rotation is thought to be due to major mergers and thus studies of prolate-rotating systems can help us better understand the hierarchical process of galaxy evolution. Dynamical studies of such galaxies are important to find their gravitational potential profile, total mass, and dark matter fraction. Recently, it has been shown in a cosmological simulation that it is possible to form a prolate-rotating dwarf galaxy following a dwarf-dwarf merger event. The simulation also shows that the unusual prolate rotation can be time enduring. In this particular example, the galaxy continued to rotate around its major axis for at least $7.4$\,Gyr (from the merger event until the end of the simulation). In this project, we use mock observations of the hydro-dynamically simulated prolate-rotating dwarf galaxy to fit various stages of its evolution with Jeans dynamical models. The Jeans models successfully fit the early oblate state before the major merger event, and also the late prolate stages of the simulated galaxy, recovering its mass distribution, velocity dispersion, and rotation profile. We also ran a prolate-rotating N-body simulation with similar properties to the cosmologically simulated galaxy, which gradually loses its angular momentum on a short time scale $\sim100$\,Myr. More tests are needed to understand why prolate rotation is time enduring in the cosmological simulation, but not in a simple N-body simulation.

\end{abstract}


\maketitle

\section{Introduction}
Dwarf galaxies are important building blocks in galaxy hierarchical formation and evolution theories, so it is essential to understand their properties. There are distinct types of dwarf galaxies, depending on mass and gas content - mainly gas-rich dwarf irregulars and gas-poor dwarf spheroidals, as well as transition types. The majority of them are dark matter (DM) dominated. However, their low stellar densities and shallow potential wells make them very sensitive to perturbations from stellar feedback, external interactions, etc.

Dynamical studies of dwarf galaxies can reveal their total mass and the mass of their sub-components (stars, gas, DM content), as well as their internal structure and kinematical properties, which is not possible from photometry alone.


On the other hand, cosmological simulations allow us to study the formation and evolution of galaxies at all scales and test real-time observations. By comparing observations and the simulated results we can better understand the physical processes in our Universe. Zoom-in simulations are used to study smaller-scale processes in individual galaxies and galaxy clusters, such as interactions, gas cooling, star formation, and feedback from supernovae and black holes.

While most galaxies have a certain amount of angular momentum, which flattens them in the direction of the rotation axis, there are rare exceptions of galaxies that rotate around their major axis, known as prolate rotation.
In the Local Group, we know of only two such cases, the And II dwarf \citep{amorisco2014} and the Phoenix dwarf \citep{kacharov2017}.
In both studies, they strongly suggest that a major merger at some point in the galaxy's evolution causes a change in the rotation axis. 
Similarly, \citet{cardona2021} discovered the presence of prolate rotation in a cosmologically simulated dwarf galaxy, using the {\sc gear} code, based on {\sc gadget-2} \citep{springel2005, revaz2012}.
The culprit, causing the flip of the rotation axis is also a major dwarf-dwarf merger with a DM mass ratio of about $1:5$.

In this study, we aim to test how well Jeans dynamical models can reproduce the mass distribution and kinematic properties of the simulated prolate-rotating dwarf.
If successful, such dynamical models can then be applied to real observational data.

\section{Jeans Anisotropic Multi-Gaussian Expansion (JAM) on the cosmological data}
Jeans equations are stellar hydrodynamic equations, derived from the collisionless Boltzmann equation, that are used to study the dynamics of equilibrium systems.
One of the advantages of the Jeans equations is that instead of dealing with the phase-space distribution function, we use velocity moments of the line-of-sight velocity distribution (LOSVD), which is directly linked to the gravitational potential and density of the system.
This method has been extensively used to study spherical and axisymmetric stellar systems under the assumption of quasi-equilibrium \citep{cappellari2016}.

In this work, we utilise the JAM code by \citet{cappellari2020} to solve the Jeans equations for the cosmologically simulated prolate rotating dwarf galaxy, discovered by \citet{cardona2021}.
We obtain best-fit Jeans dynamical models at different stages during its evolution in the cosmological simulation, which predicts the dwarf's mass distribution and compares the model results with the known properties of the simulated dwarf. 
The primary objective is to investigate whether the Jeans models can accurately describe the dynamical properties of the prolate system in a quasi-equilibrium approximation since in the cosmological simulation the prolate rotation survives for more than $7.56$\,Gyr.  

In the early stages of the galaxy's evolution, its rotation axis was the minor axis ($q=\frac{b}{a}<1$).
After a massive merger with its satellite halo, the galaxy's axis of rotation changes to the major axis. We assumed here axisymmetry around its major axis ($q=\frac{b}{a}>1$). 
We created JAM models for each of the three main stages of the galaxy's evolution: pre-merger ($z=1.58$), shortly after the merger ($z=0.58$), and at the present day ($z=0.00$).

We used the initial oblate case as a benchmark to evaluate the performance of our prolate models.

\section{Methods}
The Jeans dynamical models link the generally derived from observations stellar density and kinematics of the modelled system to its gravitational potential.

From the simulation, we know the surface brightness profile of the galaxy and perform a simple S\'ersic profile fit to obtain its central surface brightness, half-light radius, and S\'ersic index. We represent the resulting S\'ersic profile with a Multi Gaussian Expansion (MGE), which makes it easy to deproject the density profile.

We have also created Voronoi maps of the line of sight velocity ($v_{los}$) and velocity dispersion ($\sigma$) at different snapshots from the simulation. We assigned uncertainties to these values based on the local surface brightness (uncertainties increase with decreasing surface brightness).

In our dynamical model, the gravitational potential is driven by DM only. We used a generalised Navarro-Frank-White (NFW) density profile as representative of the gravitational potential to solve the Jeans equations. In this particular JAM model, we used a spherical, cored NFW profile (central density slope $\gamma=0$), which we later found to give the most promising mass profile of the model, as compared with the cosmological simulation. 

We used Markov Chain Monte Carlo (MCMC) method to get the maximum likelihood of the fitted parameters. We fit for the central density ($\rho_0$) and scale radius ($r_0$) of the NFW profile, as well as velocity anisotropy $(\beta_z= 1- \frac{\sigma_z^2}{\sigma_r^2})$ and rotation amplitude ($\kappa$) of the galaxy.

\section{Results}


We first modelled the simulated galaxy during its pre-merger stage ($z=1.58$), when it still had normal oblate rotation.
After $2$\,Gyr, due to a massive collision, the galaxy began rotating in a prolate manner. This is the stage in the simulation, at which we performed our second JAM model ($z=0.58$).
Interestingly, the galaxy also has prolate rotation at $z=0$ (the present day of the simulation), indicating that it is in a state of quasi-equilibrium.

\begin{figure}
    \centering
    \includegraphics[width=0.4\textwidth]{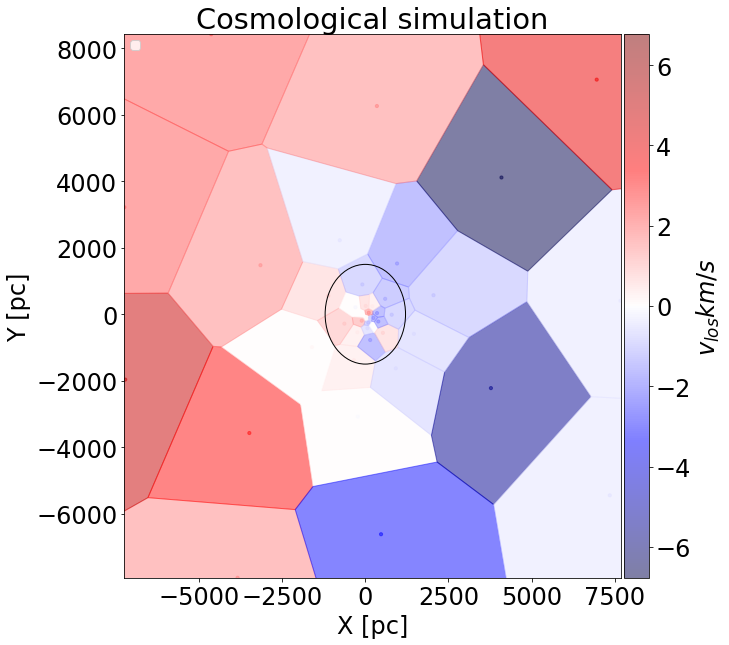}
    \includegraphics[width=0.4\textwidth]{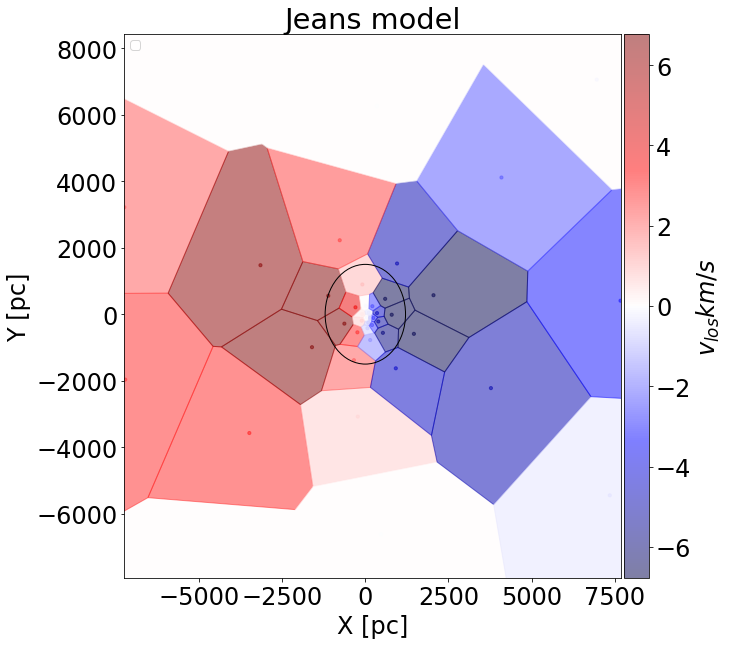}
    \\
    \includegraphics[width=0.4\textwidth]{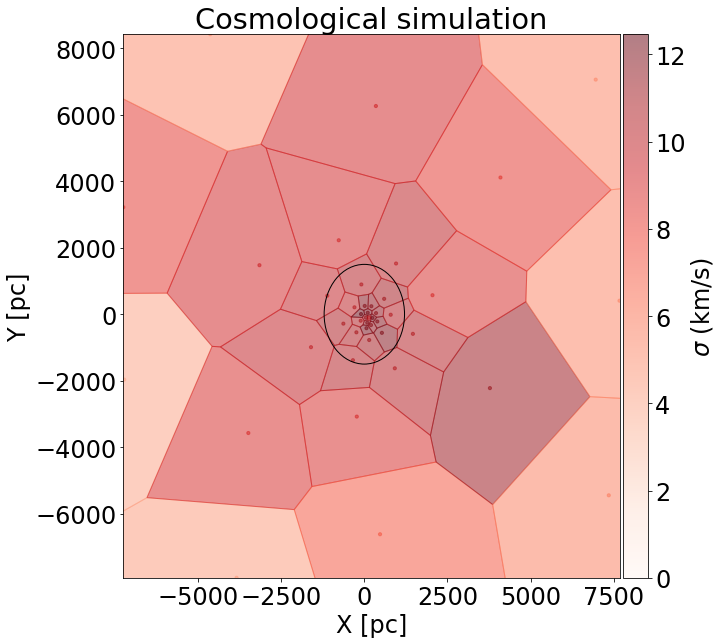} 
    \includegraphics[width=0.4\textwidth]{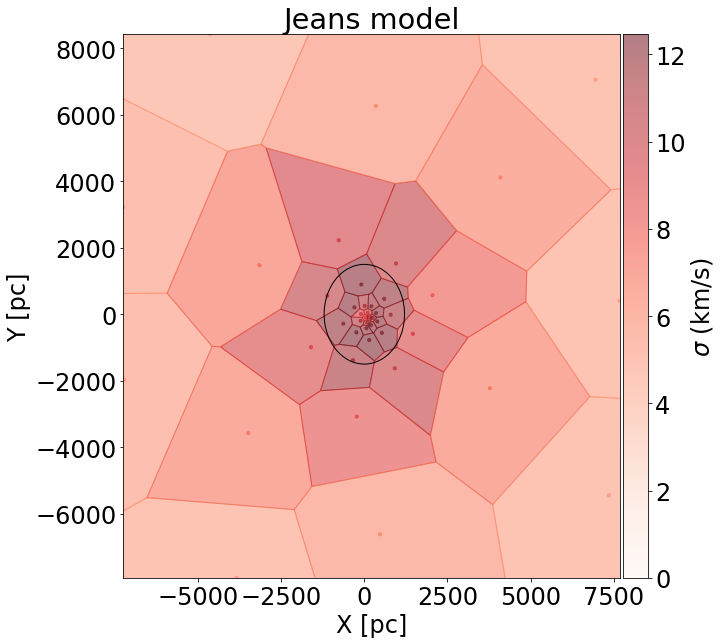}
    
    \caption{The top-left figure displays the line-of-sight velocity from the cosmological simulation of the galaxy at redshift $z=0.00$ in Voronoi bins, while the top-right figure shows the corresponding JAM model. The black ellipse indicates the half-light radius. The bottom left figure displays the velocity dispersion from the cosmological simulation, while the bottom right shows the dispersion obtained from the fitted JAM model. }
    \label{fig1}
\end{figure}

In all three cases, we fitted the line-of-sight velocity ($v_{los}$) and velocity dispersion, and we estimated the mass distribution of the simulated galaxy using the modelled parameters listed above.
An example of the fit for the last case ($z=0$) is shown in Figure \ref{fig1}.
The Voronoi maps of the mock observations of the galaxy rotation and velocity dispersion are compared to the best-fit results from the Jeans model.

The Jeans model provided a well matched line-of-sight velocity and velocity dispersion for all the stages within the half-light radius. Despite the fewer data points outside the half-light radius, we also got a relatively good fit with the data.  



\begin{figure}[]
    \centering
    \includegraphics[width=0.3\textwidth]{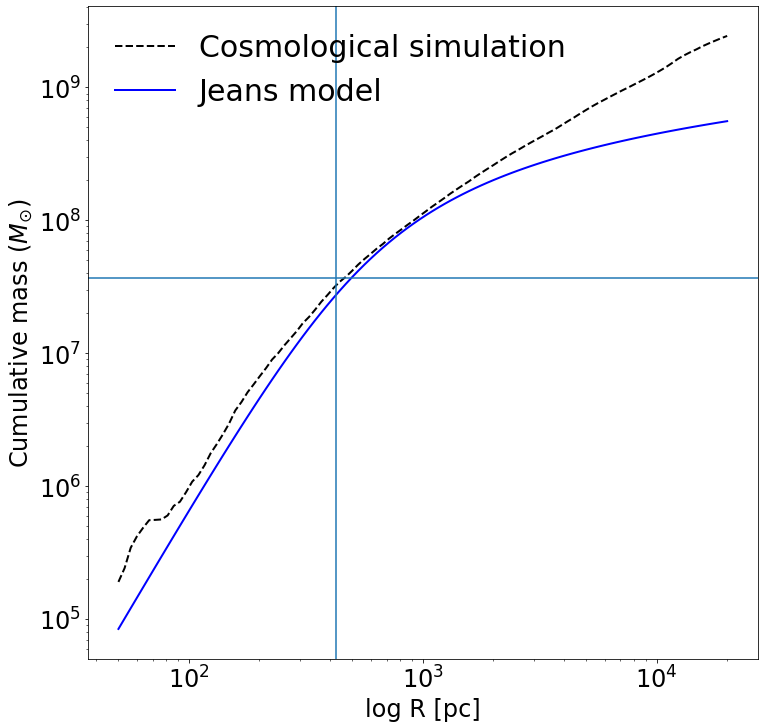}
        \includegraphics[width=0.3\textwidth]{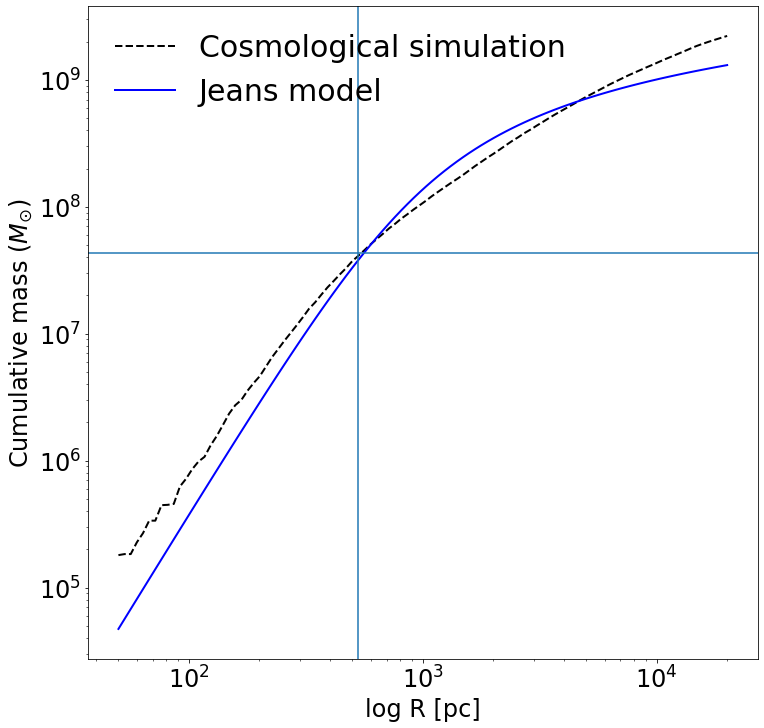}
            \includegraphics[width=0.3\textwidth]{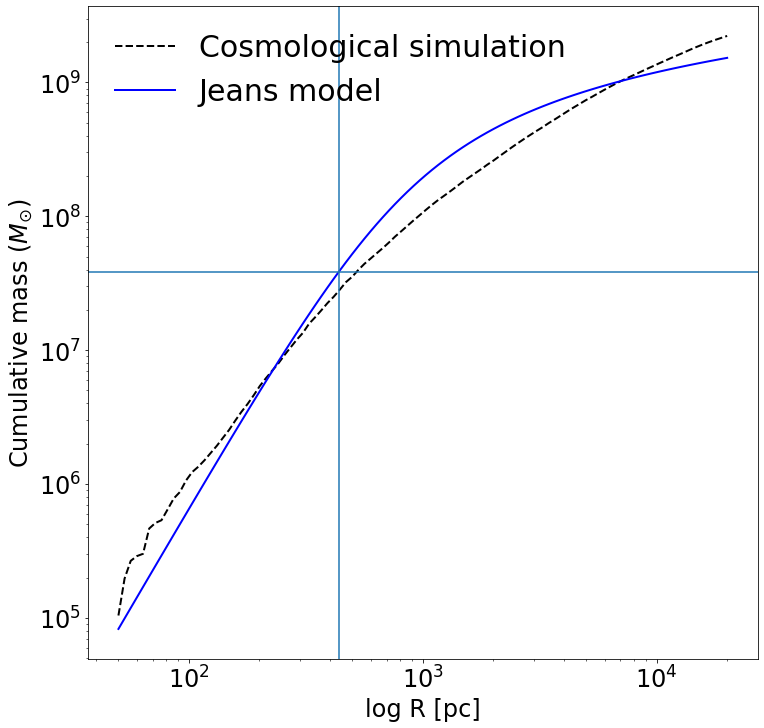}
    \caption{Best-fit mass profiles of the simulated galaxy at pre-merger ($z=1.58$, left), just after the merger ($z=0.58$, middle), and at the present time ($z=0$, right), compared to the true mass distribution from the cosmological simulation. The cross shows the virial mass estimate enclosed within the half-light radius.}
    \label{mass}
\end{figure}

The capability to determine the galaxy's mass profile is one of the main advantages of dynamical modelling. We calculated the mass for each of the three models stated above. The masses for all three models are displayed in Figure \ref{mass}, along with the mass derived from the cosmological simulation.

\section{N-body Simulation of the galaxy}
 We also checked the survivability of the galaxy's rotation using an N-body simulation.
 
 We used the inverse transform sampling method to draw positions from the best-fit de-projected 3D s\'ersic profile and initial 3D velocities from the best-fit Jeans model.
 
 We ran the N-body code assuming star particles move within an embedded spherical DM halo. The simulation was done using SWIFT \citep{swift}. Interestingly, the galaxy loses its prolate rotation after only $100$\,Myr, in contrast to its longevity in the cosmological simulation.
 More research is necessary to determine the underlying reasons.
 For example, we could try changing the shape of the DM halo to prolate as well.

\section{Conclusions}
We successfully applied the JAM code on a cosmologically simulated, prolate rotating dwarf galaxy and fitted its velocity moments. We used the early oblate, pre-merger stage as a benchmark for our project. We then modelled the prolate rotating stages shortly after the merger ($z=0.58$) and at the present time ($z=0.0$).
In all cases, we fitted a spherical NFW mass profile, as well as the amplitude of rotation and velocity anisotropy.
We then calculated the mass profile of the galaxy at different evolutionary stages and compared them to the known mass profiles from the simulation, which agree quite well.
This project confirms the viability of the Jeans method for estimating the masses of prolate rotating galaxies in the approximation of quasi-dynamical equilibrium.
Furthermore, we also did an N-body simulation but in this simple setup, the galaxy's prolate rotation dies after $100$\,Myr. Further work is needed to better understand the survivability of its prolate nature.

\section*{Acknowledgements}
We want to express our gratitude to Dr. Salvador Cardona-Barrero for providing the simulation data. We acknowledge Dr. Yves Revaz for his help to run the N-body simulation. We thank Prof. Dr. Maria-Rosa Cioni for the opportunity to conduct this research at the AIP.


\begin{thebibliography}{}
\bibitem[Amorisco et al.(2014)]{amorisco2014} Amorisco, N.~C., Evans, N.~W., \& van de Ven, G.\ 2014, Nature, 507, 335. doi:10.1038/nature12995
\bibitem[Cappellari(2016)]{cappellari2016} Cappellari, M.\ 2016, Annu. Rev. Astron. Astrophys., 54, 597. doi:10.1146/annurev-astro-082214-122432
\bibitem[Cappellari(2020)]{cappellari2020} Cappellari, M.\ 2020, MNRAS, 494, 4819. doi:10.1093/mnras/staa959
\bibitem[Cardona-Barrero et al.(2021)]{cardona2021} Cardona-Barrero, S., Battaglia, G., Di Cintio, A., et al.\ 2021, MNRAS, 505, L100. doi:10.1093/mnrasl/slab059
\bibitem[Kacharov et al.(2017)]{kacharov2017} Kacharov, N., Battaglia, G., Rejkuba, M., et al.\ 2017, MNRAS, 466, 2006. doi:10.1093/mnras/stw3188
\bibitem[Revaz \& Jablonka(2012)]{revaz2012} Revaz, Y. \& Jablonka, P.\ 2012, A\&A, 538, A82. doi:10.1051/0004-6361/201117402
\bibitem[Schaller et al.(2016)]{swift} Schaller, M., Gonnet, P., Chalk, A.~B.~G., et al.\ 2016, Proceedings of the Platform for Advanced Scientific Computing Conference, 2. doi:10.1145/2929908.2929916
\bibitem[Springel(2005)]{springel2005} Springel, V.\ 2005, MNRAS, 364, 1105. doi:10.1111/j.1365-2966.2005.09655.x


\end{thebibliography}
\end{document}